\documentclass[prb,aps,twocolumn,superscriptaddress,showpacs]{revtex4}
\usepackage{epsfig}
\usepackage{amsmath}
\usepackage{amsthm}
\usepackage{amssymb,braket}
\usepackage{verbatim}
\usepackage{bm}
\usepackage{hyperref}
\usepackage[all]{hypcap}
\usepackage{here}
\usepackage{sidecap}
\usepackage{float}

\newcommand{\bpm}{\begin{pmatrix}}
\newcommand{\epm}{\end{pmatrix}}
\newcommand{\ba}{\begin{eqnarray}}
\newcommand{\ea}{\end{eqnarray}}

\newcommand{\smk}{\text{\tiny{\emph{K}}}}

\begin{document}
\title{Microscopic mechanism for asymmetric charge distribution in\\ Rashba-type surface states and the origin of the the energy splitting scale}

\author{Beomyoung Kim}
\affiliation{Institute of Physics and Applied Physics, Yonsei University, Seoul 120-749, Korea}

\author{Panjin Kim}
\affiliation{Department of Physics and BK21 Physics Research Division, Sungkyunkwan University, Suwon 440-746, Korea}

\author{Wonsig Jung}
\affiliation{Institute of Physics and Applied Physics, Yonsei University, Seoul 120-749, Korea}

\author{Yeongkwan Kim}
\affiliation{Institute of Physics and Applied Physics, Yonsei University, Seoul 120-749, Korea}

\author{Yoonyoung Koh}
\affiliation{Institute of Physics and Applied Physics, Yonsei University, Seoul 120-749, Korea}

\author{Wonshik Kyung}
\affiliation{Institute of Physics and Applied Physics, Yonsei University, Seoul 120-749, Korea}

\author{Joonbum Park}
\affiliation{Department of Physics, Pohang University of Science and Technology, Pohang 790-784, Republic of Korea}

\author{Masaharu Matsunami}
\affiliation{UVSOR Facility, Institute for Molecular Science and Graduate University for Advanced Studies, Okazaki 444-8585, Japan}

\author{Shin-ichi Kimura}
\affiliation{UVSOR Facility, Institute for Molecular Science and Graduate University for Advanced Studies, Okazaki 444-8585, Japan}

\author{Jun Sung Kim}
\affiliation{Department of Physics, Pohang University of Science and Technology, Pohang 790-784, Republic of Korea}

\author{Jung Hoon Han}
\affiliation{Department of Physics and BK21 Physics Research Division, Sungkyunkwan University, Suwon 440-746, Korea}

\author{Changyoung Kim}
\email[Electronic address:$~~$]{changyoung@yonsei.ac.kr}
\affiliation{Institute of Physics and Applied Physics, Yonsei University, Seoul 120-749, Korea}

\begin{abstract}
Microscopic mechanism for the Rashba-type band splitting is examined in detail. We show how asymmetric charge distribution is formed when local orbital angular momentum (OAM) and crystal momentum get interlocked due to surface effects. An electrostatic energy term in the Hamiltonian appears when such OAM and crystal momentum dependent asymmetric charge distribution is placed in an electric field produced from an inversion symmetry breaking (ISB). Analysis by using an effective Hamiltonian shows that, as the atomic spin-orbit coupling (SOC) strength increases from weak to strong, originally OAM-quenched states evolve into well-defined chiral OAM states and then to total angular momentum $J$-states. In addition, the energy scale of the band splitting changes from atomic SOC energy to electrostatic energy. To confirm the validity of the model, we study OAM and spin structures of Au(111) system by using an effective Hamiltonian for the $d$-orbitals case. As for strong SOC regime, we choose Bi$_2$Te$_2$Se as a prototype system. We performed circular dichroism angle resolved photoemission spectroscopy experiments as well as first-principles calculations. We find that the effective model can explain various aspects of spin and OAM structures of the system.

\end{abstract}
\maketitle

%Introduction%

\section{Introduction}
In the conventional interpretation of the Rashba model,\cite{Rashba} the electron spin interacts with an effective magnetic field due to the electron's relativistic motion in an electric field, resulting in Zeeman splitting. The model predicts, among others, spin-degeneracy lifting with a chiral spin structure. This prediction has been experimentally verified in surface states of metals \cite{Lashell, Osterwalder, 4Ast, 4Koroteev, 4Hirahara, 5Pacile, 5Ast, 5Ast2, 5Meier} and topological insulators (TIs) \cite{Hasan} as well as interfacial states of hetero structures,\cite{Nitta} all of which possess inversion symmetry breaking (ISB).

In spite of its success in predicting the band splitting and a chiral spin structure, the Rashba model has multiple unresolved issues. First of all, the predicted energy scale of the band splitting is much smaller than the experimentally measured value (for example, about $10^{5}$ times smaller for Au(111) surface bands).\cite{Lashell} In addition, the effect can only be observed in high atomic number materials.\cite{Lashell,Osterwalder, 4Ast, 4Koroteev, 4Hirahara} Several arguments have been proposed to resolve these issues. While some groups emphasize the importance of the potential gradient along the surface normal times the charge density as a crucial factor to determine the magnitude of Rashba spin splitting,\cite{Nagano, Yaji} others argue that splitting energy comes from strong in-plane gradient of the crystal field in the surface layer.\cite{5Ast, Bihlmayer, Premper, Frantzeskakis}

Above-mentioned arguments are, however, somewhat speculative and cannot fully explain all aspects of the Rashba-related phenomena. Recently, a new interpretation for the Rashba effect has been proposed for materials with strong atomic spin-orbit coupling (SOC).\cite{SRParkPRL} The new proposal attributes the Rashba-type band splitting to formation of asymmetric charge distribution from local orbital angular momentum (OAM) and crystal momentum. The energy splitting comes from the interaction between the electrostatic field due to ISB and the asymmetric charge distribution. The relationship among the directions of asymmetric charge distribution, OAM and crystal momentum naturally points to a chiral OAM structure, and the chiral spin structure is only concomitant to the chiral OAM structure.\cite{SRParkPRL}

After the new model was proposed, it was also found experimentally and theoretically that chiral OAM structure exists in surface states of metals\cite{BYKim,JHPark,HJLee,Jakobs} even when SOC is weak. Somewhat surprisingly, it was found that the alignment(parallel or anti-parallel) between spin and OAM are band and crystal momentum dependent.\cite{BYKim} These results are explained within an effective model with a range of atomic SOC strength $\alpha$.

\begin{SCfigure*}
\centering \epsfxsize=12.5cm \epsfbox{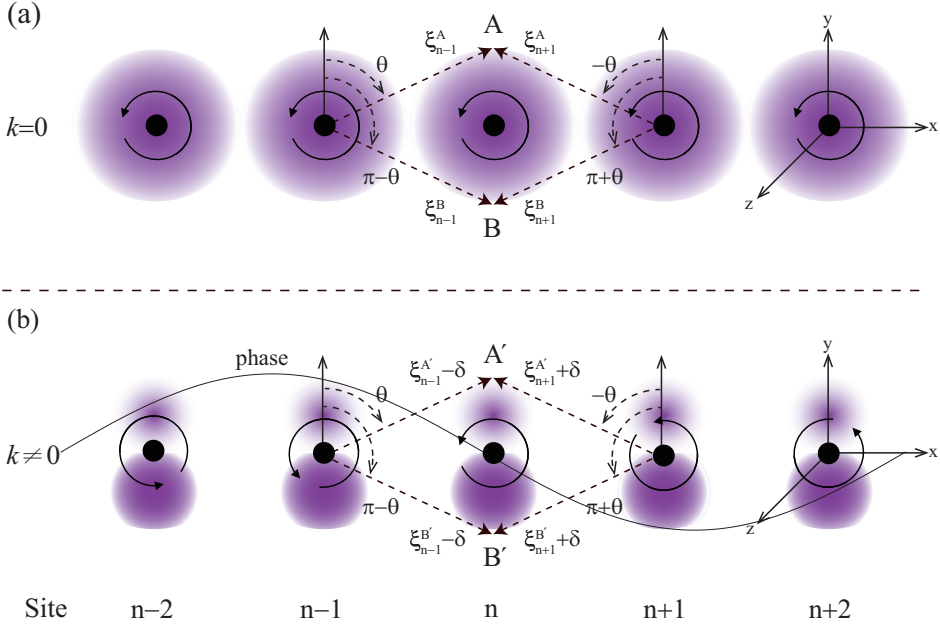} \caption{(Color online) Mechanism behind the formation of asymmetric charge distribution. Each atom has a finite OAM in the $z$-direction. The helical arrow on each atom represents the phase variation due to the OAM and the crystal momentum. $\theta$ is angle between the $y$-axis and the point of interest. $\xi^{A}_{n}$ is a phase at the point A of the orbital of the n-th atom. Also plotted in color are electron densities for (a) zero and (b) non-zero crystal momenta. $\delta$ is an additional phase due to the crystal momentum.}\label{Fig.1}
\end{SCfigure*}

The purpose of this work is to expand our earlier work.\cite{SRParkPRL,BYKim} We give detailed explanation for the mechanism behind the formation of asymmetric charge distribution, for which our earlier work was also somewhat speculative. Furthermore, we describe in detail characters of spin and OAM states for different sizes of atomic SOC parameter $\alpha$, in terms of the three major terms (crystal field, atomic SOC, and electrostatic) in the Hamiltonian. With the further understanding, we can describe the band splitting mechanism in more detail for different sizes of SOC strength. To confirm our model, we extend our effective model to $d$-orbital states and study spin and OAM structures of Au(111) surface states as a weak SOC case. As for strong SOC regime, We take Bi$_2$Te$_2$Se and perform circular dichroism angle resolved photoemission (CD-ARPES) experiments as well as first-principles calculations on the system.

\section{Experimental and DFT calculation methods}
Bi$_2$Te$_2$Se single crystals were grown using the self-flux method. The stoichiometry and high crystallinity of the single crystals were confirmed by energy dispersive spectroscopy and X-ray diffraction measurements, respectively. ARPES measurements were performed at the beamline 7U of UVSOR equipped with MB Scientific A-1 analyzer.\cite{UVSOR} Samples were cleaved $in$ $situ$ to obtain clean surfaces. Data were taken with right and left circularly polarized (RCP and LCP, respectively) $10$ eV light. Circular polarization was achieved by using a MgF$_2$ $\lambda /4$ wave plate. The total energy resolution was set to be $10$ meV at $10$ eV, and the angular resolution was $0.1^{\circ}$ which corresponds a momentum resolution of $0.0016 \AA^{-1}$. Experiments were performed at $10$ K under a base pressure better than $6.7\times 10^{-11}$ Torr.

In order to examine the spin and OAM structures of Bi$_2$Te$_2$Se in more detail, we performed first-principles density-functional theory calculations using OpenMX code\cite{openmx} within the generalized gradient approximation of Perdew, Burke and Ernzerhof.\cite{Perdew} The self-consistent norm conserving pseudopotentials of Bi, Te and Se were generated from the OpenMX database. Supercells of the slab model of Bi$_2$Te$_2$Se with a thickness ranging from 1 quintuple layer (QL) to 4 QLs were used to investigate surface states. Parameters of the supercell geometry are based on the experimental data.

\section{Modeling}
\subsection{Formation of asymmetric charge distribution}

\begin{SCfigure*}
\centering \epsfxsize=12.5cm \epsfbox{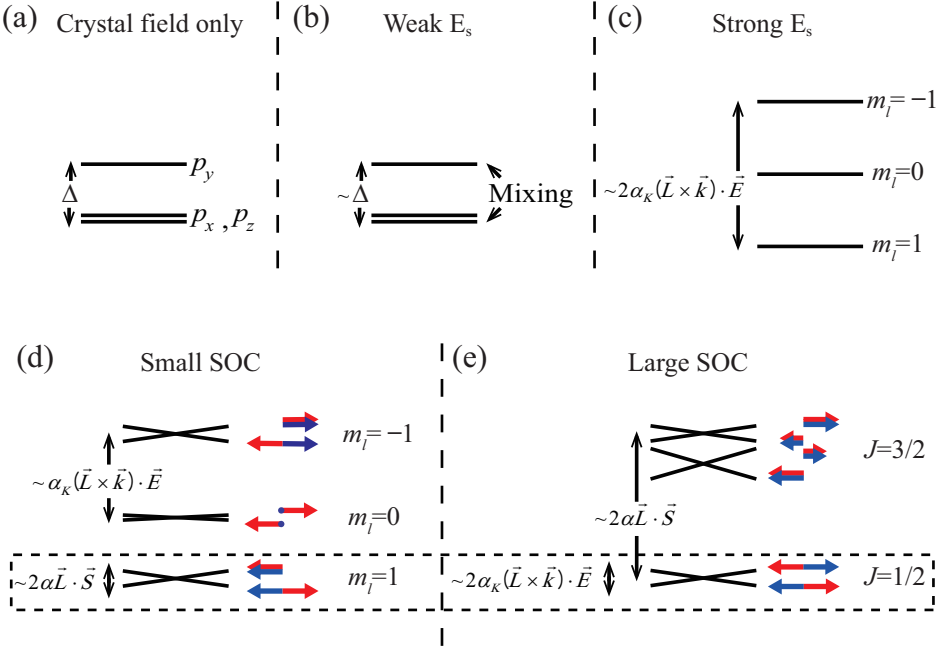} \caption{(Color online) Schematic of energy levels and angular momentum configurations of $p$-states with (a) crystal field only. Configurations when additional (b) weak $\vec{E}_S$, (c) strong $\vec{E}_S$, (d) strong $\vec{E}_S$ + weak $\vec{E}_S$, and (e) strong $\vec{E}_S$ + strong $\vec{E}_S$ are applied. $\Delta$ represents crystal field energy between in- and out-plane orbitals, and $\alpha$ the atomic SOC strength. Red (blue) arrows represent spin (OAM). For simplicity, we assume that the energy scale of $\hat{H}_{ES}$ is larger than $\Delta$. Dashed box represents Rashba-type band splitting in each case.}\label{Fig.2}
\end{SCfigure*}

The effective Hamiltonian includes three major terms as described earlier\cite{BYKim}: crystal field, atomic SOC and electrostatic terms. The electrostatic term, which stems from the interaction between asymmetric charge distribution and electrostatic field, is the most unfamiliar one. This term is derived from the Hamiltonian with SOC. It is given by
\begin{align}
    \hat{H}=\frac{\vec{p}^{2}}{2m_{e}}+V+\frac{\hbar}{4{m_{e}}^{2}{c^{2}}}(\vec{\nabla}V \times \vec{p}) \cdot \vec{\sigma}.
\end{align}
In the presence of an ISB on surface or interface, we need to consider an extra electric field in addition to the atomic field. We assume that this external field has only $z$-component and thus replace $V$ with $V_{bulk}+V_{surface}$ where $V_{surface}$ is approximately $-E_{S}z$ ($E_S$ is constant). The Hamiltonian becomes
\begin{align}
    &\hat{H}=\frac{\vec{p}^{2}}{2m_{e}}+V_{bulk}+eE_{S}z\\ \nonumber
    &\qquad+\frac{\hbar}{4{m_{e}}^{2}{c^{2}}}(\vec{\nabla}V_{bulk} \times \vec{p}) \cdot \vec{\sigma} +\frac{e \hbar E_S}{4{m_{e}}^{2}{c^{2}}} (\hat{z}\times \vec{p}) \cdot \vec{\sigma}.\cr
\end{align}
The first two terms, kinetic and potential energies in the bulk, give the usual band energy. The fourth term describes the atomic SOC, $\hat{H}_{SOC} = \alpha \vec{L} \cdot \vec{S}$, and the last term is the well-known Rashba Hamiltonian, $\hat{H}_{R} = \alpha_{R} (\vec{k} \times \vec{E_s})\cdot \vec{\sigma}$, where $\alpha _{R} = \frac{e {\hbar}^{2} E_S}{4{m_{e}}^{2}{c^{2}}}$. As mentioned above, the effect of Rashba Hamiltonian is negligible.

On the other hand, the third term is what gives the interaction between the asymmetric charge distribution and surface electrostatic field. An earlier report\cite{JHPark} provides detailed calculation for interaction between the asymmetric charge distribution and surface electrostatic field by tight-binding and first-principles calculations. According to it, the electrostatic Hamiltonian should be approximately given by $-\alpha_{\smk} (\vec{L} \times \vec{k})\cdot \vec{E_s}$ where $\alpha_{\smk}$ is a proportionality constant that increases as the overlap of atomic orbitals for a small $\vec{k}$.\cite{JHPark} Note that this term comes from a non-relativistic term in equation (2) and that the only relativistic contribution is in the atomic SOC itself.

\begin{figure*}
\centering \epsfxsize=16cm \epsfbox{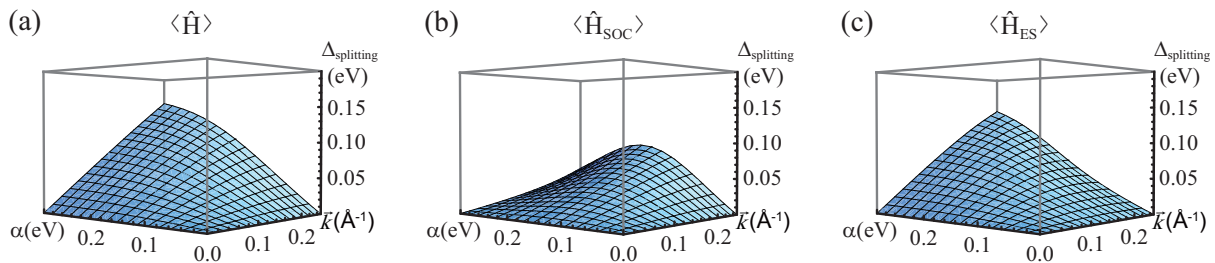} \caption{(Color online) Expectation values of different terms in the Hamiltonian of $p$-orbitals case as a function of atomic SOC strength and crystal momentum. Plotted are contributions from (a) total energy, (b) atomic SOC energy, and (c) electrostatic energy.}\label{Fig.3}
\end{figure*}

We will now show pictorially how the third term in equation (2) contributes to the formation of asymmetric charge distribution in the presence of local OAM and crystal momentum. Let us consider a Bloch state as an array of atomic orbitals with a local (atomic) OAM,
\begin{align}
    \psi_{\vec{k}}(\vec{r}) = \frac{1}{\sqrt{N}}\sum_{\nu} e^{\imath \vec{k} \cdot \vec{R}_{\nu}} \phi(\vec{r}-\vec{R}_{\nu}),
\end{align}
where $N$ is the number of lattice sites.  The Bloch state is depicted in Fig. 1 with the OAM in the $z$-direction (out of the page). The eigenvalue of $L_z$ operator is $m\hbar$, $L_{z}\phi(\vec{r}-\vec{R}_{\nu}) = m\hbar\phi(\vec{r}-\vec{R}_{\nu})$. Fig. 1(a) depicts a state with zero crystal momentum. We investigate phases of three neighboring atomic orbitals. When $\theta$ is the angle between the $y$-axis and point A, the phase $\xi$ is given by $\xi=-m\theta$. The phases of $(n-1)$, $n$ and $(n+1)$-th orbitals at point $A$ are $\xi^{A}_{n-1}=-m\theta$, $\xi^{A}_{n}=0$, $\xi^{A}_{n+1}=m\theta$, respectively. Meanwhile, the phases are $\xi^{B}_{n-1}=-m(\pi-\theta)$, $\xi^{B}_{n}=-m\pi$ and $\xi^{B}_{n+1}=-m(\pi+\theta)$ at point $B$. One can see that there is no difference in the electron density at the points $A$ and $B$ because squares of the exponentials of phase differences are the same for the two points,
\begin{small}
\begin{align}
|e^{i\xi^{A}_{n-1}}+Ce^{i\xi^{A}_{n}}+e^{i\xi^{A}_{n+1}}|^2=|e^{i\xi^{B}_{n-1}}+Ce^{i\xi^{B}_{n}}+e^{i\xi^{B}_{n+1}}|^2,
\end{align}
\end{small}where constant $C$ is proportional to distance. The situation is different for non-zero crystal momentum shown in Fig. 1(b). In this case, an additional phase $\delta$ arising from the Bloch momentum is added for the next atomic orbital as shown in the figure. Unlike the zero momentum case in Fig. 1(a), squares of the the exponentials of phase differences are different for $A^{\prime}$ and $B^{\prime}$. Consequently, the electron density becomes asymmetric (less phase difference and thus higher electron density for point $B^{\prime}$). One important aspect of our argument relating density asymmetry to OAM is that it comes from a non-relativistic term in the Hamiltonian and thus should occur even when the SOC is weak.\cite{BYKim,JHPark}

From our argument it is clear that the charge imbalance is a consequence of the phase differences induced by the OAM. Without OAM, the phases at symmetric points such as A and B will always be such that the same squared modulus (i.e. the density) will result. It may be deduced that the asymmetric charge distribution should be (at least approximately) proportional to $\vec{L}$ itself. With the direction of the OAM (taken to be along the $z$-direction) and the crystal momentum (taken to be along the $x$-direction), we find that the induced dipole moment is along the $y$-direction. With the proportionality of the induced dipole moment to both the non-zero OAM and finite $\vec{k}$, we conclude that the dipole moment must have the functional form $\vec{p}=\alpha_{\smk} (\vec{L} \times \vec{k})$.\cite{BYKim,JHPark} Then the Hamiltonian is given approximately by
\begin{small}
\begin{align}
\hat{H}_{ES}=-\vec{p} \cdot \vec{E_s} = -\alpha_{\smk} (\vec{L} \times \vec{k})\cdot \vec{E_s} = -\alpha_{\smk} (\vec{k} \times \vec{E_s})\cdot \vec{L}.
\end{align}
\end{small}
It is similar to the well-known Rashba Hamiltonian with the spin operator replaced by the OAM operator but can account for the energy splitting scale ($\vec{p} \cdot \vec{E_s}\sim e\cdot{\AA} \times V/{\AA}\sim eV$).\cite{JHPark,SRParkPRL,BYKim}

Our argument and the conclusion regarding the form of the electrostatic energy derived in equation (5) remains valid when the crystal momentum is sufficiently close to any time reversal invariant momentum (TRIM) points. Since $\delta$ ought to be $n\pi$ ($n$ is an integer) at TRIM points, interference effects vanish leading to equal charge densities at point $A^{\prime}$ and $B^{\prime}$.  Small deviation of the crystal momentum $\vec{k}$ from the TRIM vector $\vec{K}$ then results in the same interference-induced density imbalance already discussed for $\vec{k}$ near 0. Therefore, we may replace the crystal momentum $\vec{k}$ by $\vec{K}_{TRIM}+\vec{k}^{\prime}$.

\subsection{Origin of the energy splitting scale}

The actual electronic structure is determined by the competition among the new electrostatic Hamiltonian $\hat{H}_{ES} = -\alpha_{\smk} (\vec{L} \times \vec{k})\cdot \vec{E_s}$, atomic SOC $\hat{H}_{SOC}$ (fourth term in equation (2)) and the crystal field $\hat{H}_{CF}$ energies.\cite{BYKim} If either $\hat{H}_{ES}$ or $\hat{H}_{SOC}$ is significant (relative to $\hat{H}_{CF}$), the system tries to lower the energy by having non-zero OAM. In fact, previous experimental and theoretical studies have shown existence of OAM in both strong and weak SOC regimes.\cite{BYKim,JHPark,HJLee} In the strong SOC regime, the system prefers to have non-zero OAM due to $\hat{H}_{SOC}$, while it is $\hat{H}_{ES}$ that induces non-zero OAM in the weak SOC regime. In our previous work,\cite{BYKim} Au(111) surface states were presented as an example of a weak SOC system. It is a weak SOC system because, while inner and outer band spins are anti-parallel to each other, OAM directions are parallel (that is, spin and OAM do not form $J$-states). Even though energy states were obtained for weak SOC regime by using an effective Hamiltonian, the mechanism behind the energy splitting was not thoroughly described.\cite{BYKim} Therefore, we try to describe below how the characters of states evolve as the atomic SOC strength $\alpha$ varies.

Figure 2 shows schematic of energy levels and angular momentum configurations of $p$-states as a function of atomic SOC strength $\alpha$. In the case of no SOC and no ISB shown in Fig. 2(a), there are spin-degenerate states with no OAM. The energy splitting is solely determined by the crystal field energy $\Delta$. When an additional electric field $\vec{E}_S$ appears due to an ISB, OAM is formed and the states eventually evolve into OAM based $m = -1, 0, 1$ states with the energy scale determined by $\hat{H}_{ES}= -\alpha_{\smk} (\vec{L} \times \vec{k})\cdot \vec{E_s}$ (Fig. 2(c)). When a small SOC is present, the spin-degeneracy is lifted and states split further into spin non-degenerate states (Fig. 2(d)), with the energy splitting scale determined by $\hat{H}_{SOC} = \alpha \vec{L} \cdot \vec{S}$. Note that spin and OAM align differently for the two spin splitting bottom bands. As for the strong SOC case illustrated in Fig. 2(e), total angular momentum $J$-states become a better representation. Spin and OAM remain anti-parallel ($J=1/2$ states) or parallel ($J=3/2$ states).\cite{SRParkPRL} The larger energy scale is the energy difference between $J=1/2$ and $3/2$ states, and is determined by $\hat{H}_{SOC}$. Meanwhile, the smaller energy scale of the split bands ($m_J = \pm 1/2$ states) is determined by $\hat{H}_{ES}$.

The validity of the above description can be checked with the expectation value of each contribution to the total energy splitting in the dashed box in Fig. 2. Plotted in Fig. 3 are the total $\hat{H}$, atomic SOC $\hat{H}_{SOC}$ and electrostatic $\hat{H}_{ES}$ energies as a function of atomic SOC strength and crystal momentum.\cite{BYKim} The total energy in Fig. 3(a) shows that for a given $\alpha$, the energy splitting is approximately proportional to the crystal momentum $\vec{k}$ as expected. Looking at the contribution from $\hat{H}_{SOC}$ in Fig. 3(b), for a fixed $\vec{k}$, we find it initially increases but then decreases to an insignificant number as $\alpha$ increases. This means that $\hat{H}_{SOC}$ determines the energy splitting for a small SOC case but it is insignificant for a large SOC case. The contribution from $\hat{H}_{ES}$ plotted in Fig. 3(c) shows that it is $\hat{H}_{ES}$ that determines the energy splitting in the large SOC regime. These results are very much consistent with the above description. In spite of the change in the role, the energy splitting remains to be linear in crystal momentum $\vec{k}$. In the weak SOC regime, the OAM size increases linearly with $\vec{k}$. Therefore, the energy splitting $\sim 2 \alpha \vec{L} \cdot \vec{S}$ is also linearly dependent on $\vec{k}$. On the other hand, OAM is fully polarized in the strong SOC regime and the energy splitting $-\alpha_{\smk} (\vec{L} \times \vec{k})\cdot \vec{E_s}$ is again linearly dependent on $\vec{k}$. This makes the energy splitting always linearly proportional to $\vec{k}$, independent of the size of $\alpha$.

\section{Application of the model to real systems}
\subsection{Au(111), weak SOC case}

We now wish to apply the effective Hamiltonian to real systems with weak and strong SOC strength $\alpha$. Even though we have discussed evolution of electronic structure as a function of $\alpha$ in our previous work,\cite{BYKim} the effective Hamiltonian was not applied to real systems. In fact, while $d$-orbitals were found to play the major role in the band splitting in Au(111) states (a weak SOC system),\cite{BYKim,HJLee} only $p$-orbitals were considered in the effective Hamiltonian for simplicity.\cite{BYKim} Here we expand our previous work and construct an effective Hamiltonian for $d$-states in the form of $10 \times 10$ matrix, similar to the $p$-orbital case of $6 \times 6$ matrix.\cite{BYKim} We take sample normal direction as y-axis and consider 4 terms in the Hamiltonian: kinetic energy  $\hat{H}_{K}$, crystal field  $\hat{H}_{CF}$, atomic SOC  $\hat{H}_{SOC}$, and the electrostatic energy $\hat{H}_{ES}$. For  $\hat{H}_{K}$, we simply add a $C{k}^{2}$ term to account for the parabolic bands of Au(111) surface states. $\hat{H}_{CF}$ is energy splitting between in-plane and out-of-plane orbitals. For $d$-orbitals, there are five states that have spin-degeneracy. Among them, ${d}_{{r}^{2}-{z}^{2}}$ and $d_{zx}$ orbitals have the lowest energy of 0. $d_{xy}$ and $d_{yz}$ have $\Delta_1$, and ${d}_{{x}^{2}-{y}^{2}}$ has $\Delta_2$ ($\Delta_1 < \Delta_2$). While $\hat{H}_{SOC}$ is simply $\alpha \vec{L} \cdot \vec{S}$, $\hat{H}_{ES}$ comes from interaction of the surface electric field and asymmetric charge distribution. The total Hamiltonian is estimated in the basis of ${Y}_{2}^{2}\uparrow$, ${Y}_{2}^{2}\downarrow$, ${Y}_{2}^{1}\uparrow$, ${Y}_{2}^{1}\downarrow$, ${Y}_{2}^{0}\uparrow$, ${Y}_{2}^{0}\downarrow$, ${Y}_{2}^{-1}\uparrow$, ${Y}_{2}^{-1}\downarrow$, ${Y}_{2}^{-2}\uparrow$, ${Y}_{2}^{-2}\downarrow$. We assume the momentum is along the $x$-axis without loss of generality. The result is given by
\begin{tiny}
\begin{widetext}
{\begin{align}\label{matrix}\hsize 16cm
    &\hat{H}= Ck^2\hat{I} + \nonumber\\
    &\left(
      \begin{array}{cccccccccc}
        -2A +\alpha + \frac{\Delta_1 +\Delta_2}{2}& 0 & 0 & 0 & 0 & 0 & 0 & 0 & \frac{\Delta_1 +\Delta_2}{2} & 0 \\
        0 & -2A -\alpha + \frac{\Delta_1 +\Delta_2}{2} & \alpha & 0 & 0 & 0 & 0 & 0 & 0 & \frac{\Delta_1 +\Delta_2}{2} \\
        0 & \alpha & -A +\frac{\alpha}{2} + \frac{\Delta_1}{2} & 0 & 0 & 0 & \frac{\Delta_1}{2} & 0 & 0 & 0 \\
        0 &  0 & 0 & -A -\frac{\alpha}{2} + \frac{\Delta_1}{2} & \sqrt{\frac{3}{2}}\alpha & 0 & 0 & \frac{\Delta_1}{2} & 0 & 0\\
        0 &  0 & 0 & \sqrt{\frac{3}{2}}\alpha & 0 & 0 & 0 & 0  & 0 & 0  \\
        0 &  0 & 0 & 0 & 0 & 0 & \sqrt{\frac{3}{2}}\alpha & 0  & 0 & 0  \\
        0 &  0 & \frac{\Delta_1}{2} & 0 & 0 & \sqrt{\frac{3}{2}}\alpha & A -\frac{\alpha}{2} + \frac{\Delta_1}{2} & 0  & 0 & 0  \\
        0 &  0 & 0 & \frac{\Delta_1}{2} & 0 & 0 & 0 & A +\frac{\alpha}{2} + \frac{\Delta_1}{2}  & \alpha & 0  \\
        \frac{-\Delta_1 +\Delta_2}{2} &  0 & 0 & 0 & 0 & 0 & 0 & \alpha &  2A -\alpha + \frac{\Delta_1 +\Delta_2}{2} & 0  \\
        0 &  \frac{-\Delta_1 +\Delta_2}{2} & 0 & 0 & 0 & 0 & 0 & 0  & 0 & 2A +\alpha + \frac{\Delta_1 +\Delta_2}{2}  \\
      \end{array} \nonumber
    \right)\nonumber\\
    \nonumber\\
\end{align}}
\end{widetext}
\end{tiny}
where $A\equiv\alpha_{\smk} k_x E_s$. Band, spin and OAM structures can be obtained by diagonalizing the matrix. We adjusted C, $\Delta_1$, $\Delta_2$ and $\alpha_{\smk}$ until calculated bands match experimentally measured dispersions.

Fig. 4(a) plots the resulting band structure along with the direction of spin (red symbol) and OAM (blue symbol). More details on the spin and OAM structures are shown in Fig. 4(c) and 4(d). The OAM in the inner and outer bands are parallel to each other while spins are in opposite directions. In addition, the magnitude of the inner band OAM (represented by the lengths of the arrows in the figure) is larger than that of the outer band. These results are consistent with previous CD ARPES and band calculation results.\cite{BYKim,HJLee} We also find that the magnitude of OAM increases linearly with the crystal momentum.

We find that the spin and OAM alignment is opposite for inner and outer bands (that is, parallel for inner and anti-parallel for outer). This fact puts Au(111) states in the weak SOC regime. As for the energetics, we see that the outer band which has anti-parallel spin and OAM alignment has a lower energy than the inner band with parallel spin and OAM. It suggests that the energy difference between the two bands comes from $\hat{H}_{SOC}$, $i. e.$, the energy difference comes from the difference in the alignment between spin and OAM.

\begin{figure}[H]
\centering \epsfxsize=8cm \epsfbox{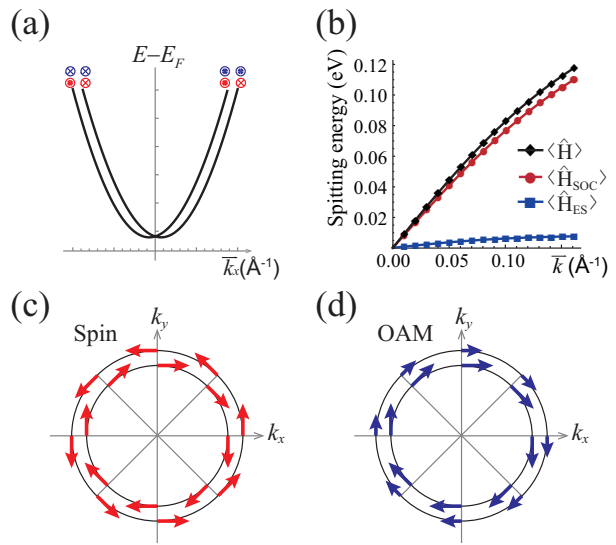} \caption{(Color online) (a) Au(111) surface state band structure obtained from the $d$-state effective Hamiltonian. Blue (red) symbol marks the OAM (spin) direction. (b) Contributions from various terms in the Hamiltonian as a function of momentum. Plotted in (c) and (d) are spin and OAM textures, respectively, on a constant energy surface.}\label{Fig.4}
\end{figure}

To confirm this fact, we calculate expectation values of $\hat{H}$, $\hat{H}_{SOC}$ and $\hat{H}_{ES}$ as a function of the crystal momentum and plot them in Fig. 4(b). First of all, all the energies increase approximately linearly with the momentum. In addition, major portion of the total energy comes from $\hat{H}_{SOC}$, confirming that $\hat{H}_{SOC}$ indeed determines the energy splitting in the weak SOC regime.

\subsection{Bi$_2$Te$_2$Se, strong SOC case}

\begin{figure}[H]
\centering \epsfxsize=8cm \epsfbox{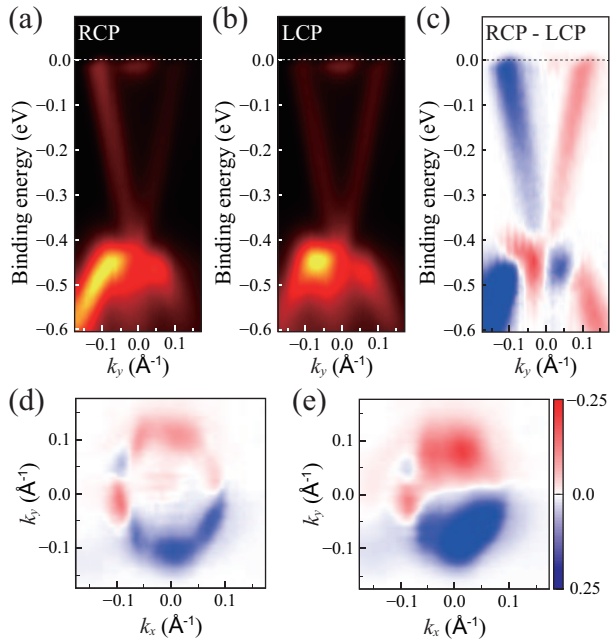} \caption{(Color online) ARPES results along the $\Gamma$-$M$ direction from Bi$_2$Te$_2$Se surface states using (a) RCP and (b) LCP light. (c) Circular dichroism data CD = RCP $-$ LCP. (d) Fermi surface CD data. (e) Cumulative CD data from the Dirac point to Fermi level.}\label{Fig.5}
\end{figure}

Finally, we move to the strong SOC regime. TIs are generally materials with very strong SOC and therefore belong to the strong SOC regime. Among various TIs, we choose Bi$_2$Te$_2$Se for the discussion. Since the relevant orbitals are Bi $p$-orbitals, we use the previously obtained effective Hamiltonian of $p$-orbitals.\cite{BYKim} As discussed in our previous study, spin and OAM are always anti-parallel to each other (for $J=1/2$) due to the strong atomic SOC and the magnitude of OAM is somewhat independent of the electron momentum $\vec{k}$.\cite{SRParkPRL,BYKim}

To confirm the result of our model in strong SOC region, we performed CD ARPES experiments and DFT calculations on Bi$_2$Te$_2$Se and investigated OAM and spin configurations. Fig. 5(a), 5(b) and 5(c) show CD-ARPES along the $\Gamma$-$M$ direction. The binding energy of the Dirac point at the $\Gamma$ point is E($\Gamma$) = 407 meV while the Fermi momentum $k_F$ = 0.097$\AA^{-1}$. These values are consistent with the published data.\cite{Arakane} RCP and LCP data in Fig. 5(a) and 5(b) show similar features except some difference in the intensity profile. Fig. 5(c) plots the CD ARPES data CD = RCP $-$ LCP. The CD ARPES data changes from red($-$) to blue($+$) as $k_y$ increases. Note that the color within one side of the band does not change much, indicating the magnitude of OAM remains approximately constant.

We also plot the Fermi surface data in Fig. 5(d). The CD pattern is roughly consistent with the chiral OAM structure in Bi$_2$Se$_3$.\cite{SRPark} Upon a closer inspection, one can see there is non-sinusoidal variation which is similar to the Bi$_2$Te$_3$ case. We therefore attribute the detailed variation to the hexagonal warping term.\cite{WSJung} To check the size of OAM as a function of the crystal momentum, we plot the CD data between the Dirac point and Fermi energy in Fig. 5(e). Uniform chiral OAM structure near the $\Gamma$-point confirms that OAM is fully polarized and its magnitude is independent of the crystal momentum. As Bi$_2$Te$_2$Se belongs to the strong SOC regime, the dominant mechanism behind the Rashba-type band splitting comes from the electrostatic term $-\alpha_{\smk} (\vec{L} \times \vec{k})\cdot \vec{E_s}$.

\begin{figure}
\centering \epsfxsize=8cm \epsfbox{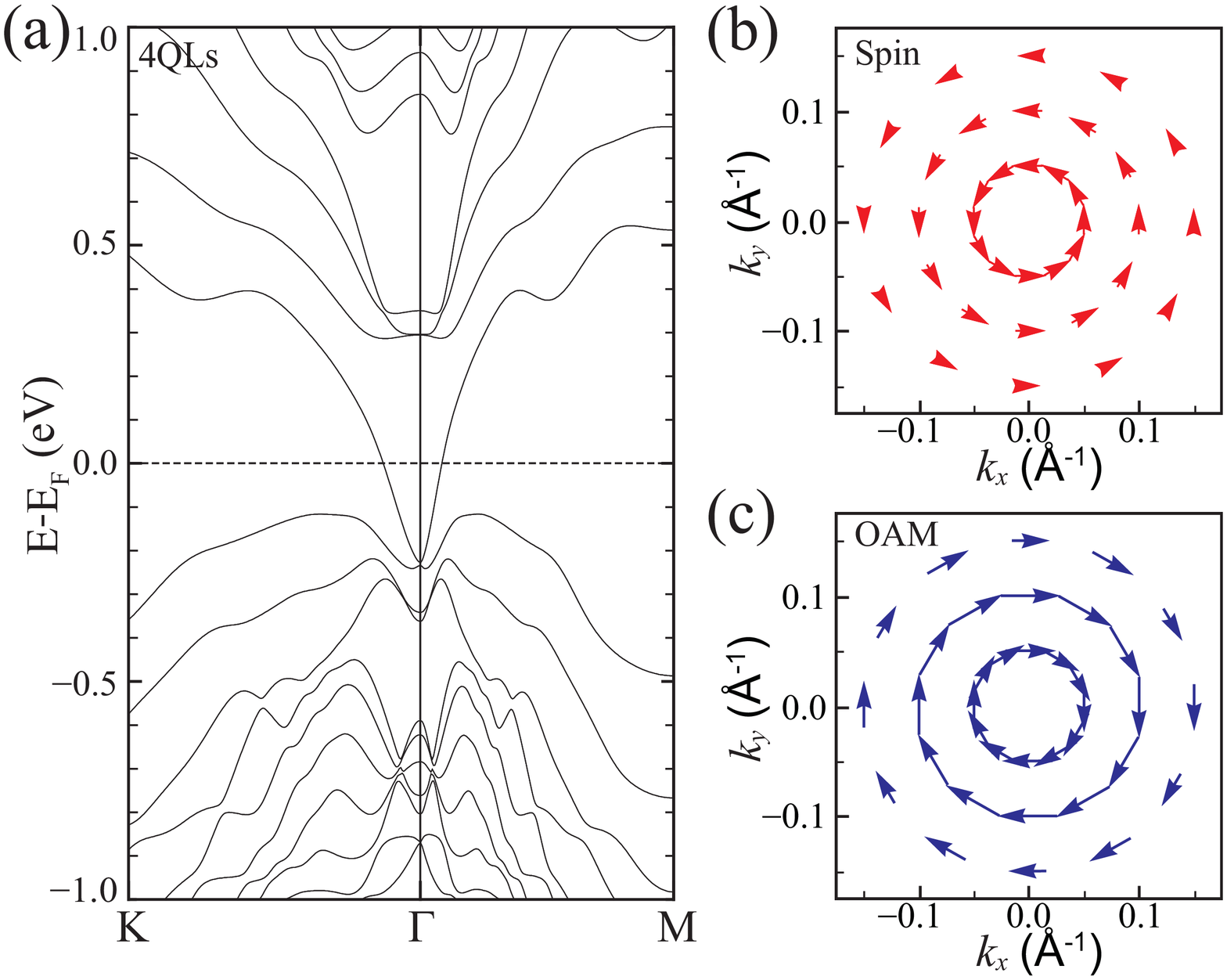} \caption{(Color online) DFT calculation result from 4QLs of Bi$_2$Te$_2$Se. (a) Band structure along the $K$-$\Gamma$-$M$ direction. In-plane components of (b) spin and (c) OAM from surface states at the Fermi energy. $\Gamma$-$K$ direction is the $k_x$-direction.}\label{Fig.6}
\end{figure}

We also performed DFT calculation to study the spin and OAM structures. Fig. 6 shows the results of DFT band calculation on Bi$_2$Te$_2$Se. We find that there is a large gap for less than 3 quintuple layers (QLs) at the $\Gamma$-point. The gap reduces to 7.4 meV for 4QLs and surface states are well separated. We plot the band structure of 4QLs in Fig. 6(a). This result is consistent with the ARPES data in Fig. 5 as well as previous calculation results.\cite{Dai,Yang}

Plotted in Fig. 6(b) and 6(c) are in-plane components of spin and OAM of surface states at $E_F$. Note that surface states at $E_F$ are from the upper Dirac cone which corresponds to the `inner band' of the Rashba-type splitting. The overall spin and OAM textures are similar to those for Bi$_2$Se$_3$.\cite{SRPark} An important aspect of the structure is that spin and OAM are anti-parallel, unlike the Au(111) case, despite that they are in the inner band. This shows that spin and OAM remain anti-parallel in all bands in the strong SOC regime. As for the magnitude of OAM, there is a small decrease in the in-plane component across some $k$-point due to formation of the out-of-plane component $k_z$ due to the warping effect as found in Bi$_2$Se$_3$ case.\cite{WSJung} Overall, OAM remains almost fully polarized. These observations put Bi$_2$Te$_2$Se in the strong SOC regime and show that the band splitting energy comes mostly from the electrostatic term $-\alpha_{\smk} (\vec{L} \times \vec{k})\cdot \vec{E_s}$.

\section{Summary}
%summary
The mechanism for Rashba-type band splitting is thoroughly examined. We first describe in detail the formation of asymmetric charge distribution that appear when OAM and crystal momentum of a tight binding state are interlocked. To better understand the mechanism, we vary the atomic SOC strength from a small value to a large value, and investigate various aspects of the states. In the weak SOC regime, orbital angular momentum $L$-states are a proper description of the eigenstates and the spin energy is dominantly described by the atomic SOC term. On the other hand, in the case of strong SOC, eigenstates are close to $J$-states and the energy is mostly determined by the electrostatic term. We extend our effective Hamiltonian to $d$-states and apply it to the Au(111) case. Finally, we performed CD ARPES experiment and DFT calculation on Bi$_2$Te$_2$Se to study spin and OAM structures. We find our effective Hamiltonian can consistently describe spin and OAM structures of the weak (Au(111) surface states) and strong (Bi$_2$Te$_2$Se) SOC cases.

\section*{ACKNOWLEDGEMENT}
The authors would like to thank Suk-young Park, Joon-Suh Park, Jinhee Han and Shoresh Soltani for helpful discussions. This work is supported by NRF (Contract No. 20100018092) and Global Research Laboratory (2011-00329) through the National Research Foundation of Korea (NRF) funded by the Ministry of Science, ICT (Information and Communication Technologies) and Future Planning. The DFT calculation is supported by a NRF grant funded by the Korean government (NRF-2012)-Global Ph.D. Fellowship Program. The work at POSTECH is supported by the Mid-Career Researcher Program (No. 2012-013838) and SRC Center for Topological Matter (No. 2011-0030785) through NRF, Korea. Part of this work was performed by the Use-of-UVSOR Facility Program (BL7U, 2012) of the Institute for Molecular Science.

\end{document}